\documentclass{iau}
 \usepackage{graphicx} 

\title[Neutron stars: history of the magnetic field decay] 
{Neutron stars: history of the magnetic field decay} 

\author[A. P. Igoshev \& A. F. Kholtygin]  
{Andrei P. Igoshev
 \and Alexander F. Kholtygin}

\affiliation{Institute of Astronomy, Saint Petersburg State University \\ email: {\tt ignotur@gmail.com}}

\pubyear{2012}
\volume{291}  
\jname{\mbox{Neutron Stars and Pulsars: Challenges and Opportunities after 80 years}}
\editors{J. van Leeuwen, ed.} 
\begin{document}

\maketitle

\begin{abstract}
  
Using the data of the ATNF pulsar catalog we study the relation
connected the real age $t$ of young neutron stars (NS) and their
spin-down age $\tau$.  We suppose that this relation is independent
from both initial period of the NS and its initial surface magnetic
field, and that the laws of the surface magnetic field decay
are similar for all NSs in the Milky Way. We further assume that the birth-rate
of pulsars was constant during at least last 200 million years. With
these assumptions we were able to restore the history of the magnetic
field decay for the galactic NSs.  We reconstruct the universal
function $f(t)=B(t)/B_0$, where $B_0$ is the initial magnetic field
and $B(t)$ is the magnetic field of NS at the age $t$. 
The function $f(t)$ can be fitted by a power law with power
index $\alpha = -1.17$.

\end{abstract}

\firstsection 
\section{Introduction}
Magnetic fields determine observational manifestations of neutron
stars. Soon after the discovery of the first radio pulsars, the decay of their 
magnetic field was supposed by~\cite{Ostriker69}. Therefore comprehension of magnetic fields behavior
aids not only in more accurately finding real pulsar ages, but also helps us understand the physical processes inside isolated neutron stars.

First of all we carefully consider behavior of spin-down age of radio pulsar:
\begin{equation}
\tau=\frac{P}{\dot P} \, .
\label{eq1}
\end{equation} 
Here $P$ is period of pulsar and $\dot P$ is its derivative. It appears that $\tau$ is highly sensitive to changes in
magnetic fields of NS. We find that it is possible to introduce an additional real pulsar age estimation based on 
the counts of pulsars with fixed spin-down age $\tau$. If we suppose
that the law governing magnetic-field decay is identical for all 
pulsars, and can be present by the formula $B(t)=B_0 f(t)$, then it is possible to restore the dimensionless factor $f(t)$ 
irrespectively of initial pulsar period distribution and initial distribution of their magnetic fields. 
This factor can be expressed as 
\begin{equation}
%
f(t)=\exp\left(\int_0^t \frac{dt'}{2\tau (t')}\right) 
\left/ \sqrt{\tau(t)} \right.
\label{f_expr}
\end{equation}
The algorithm and detailed description of our method of restoring the function $f(t)$ is described in~\cite{Igoshev2012}. 
Here we discuss the effects of the observational selection.

\section{Selection effects}

Results of restoring the function $f(t)$ by our method might be distorted by the significant observational selection effects. In fact,
we observe only a small part of all galactic pulsars. And we can not be sure exactly that this small part correctly
reproduces characteristics of the whole galactic ensemble of the NSs. Indeed, selection effects are able to hide some group of 
pulsars from our attention. In this case the selection effects can change an average and a variance of the parameters 
of the whole ensemble of the NS distribution. 

\begin{figure}[ht!]
\begin{center}
\includegraphics[width=4.8in]{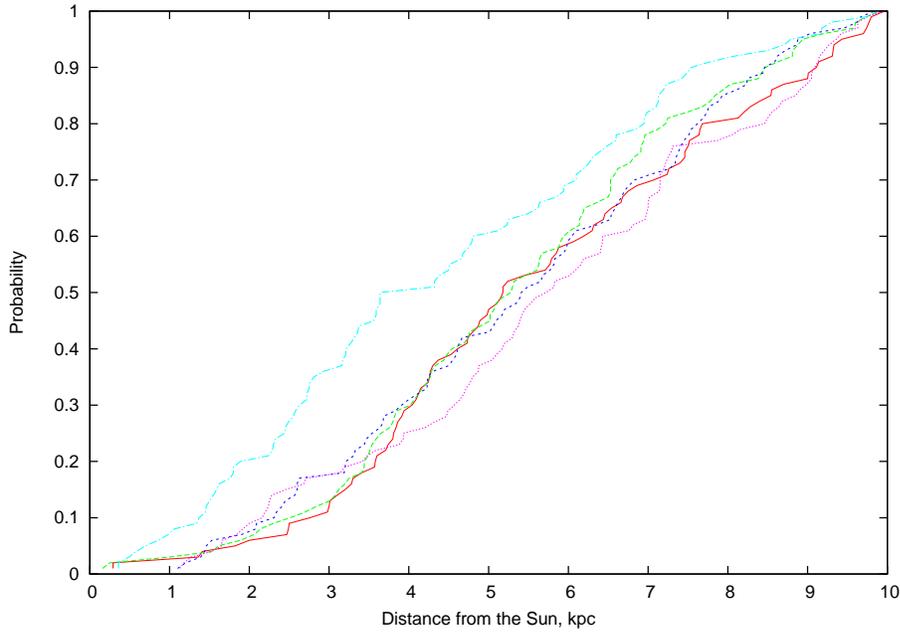}
\end{center}
    \caption{\small Cumulative distribution of the pulsars over distances from the Sun. The electronic version of 
                    article contains color figures. Here red solid line corresponds to ages $\tau\in[700, 1.2\cdot10^5]$ years, 
                    green dashed line --- to $\tau\in[1.2\cdot 10^5, 4.5\cdot 10^5]$ years, 
                    blue short-dashed line --- to $\tau\in[4.5\cdot 10^5, 9.6\cdot 10^5]$ years, 
                    violet dotted line --- to $\tau\in[9.6\cdot 10^5, 1.6\cdot 10^6]$ years
                    and sky blue dashed and dotted line - to 
                    $\tau\in[4\cdot 10^6, 6\cdot 10^6]$ years. Each interval of ages includes about of 100 pulsars. 
            }
\label{obser_select}
  \end{figure}

On the other hand it is difficult to estimate this selection effects theoretically because of our poor understanding
of the radiation mechanism for pulsars. Therefore, we will carefully analyze the observational data. We select the isolated radio-pulsars
excluding the millisecond pulsars from the ATNF pulsar catalog by~\cite{manchester05}\footnote{http://www.atnf.csiro.au/research/pulsar/psrcat/}. 
We divide them into five groups, where the spin-down age $\tau$ is expressed in years: $\tau\in [700-1.2\cdot 10^5]$,  
$\tau\in [1.2\cdot 10^5-4.5\cdot 10^5]$, $\tau\in [4.5\cdot 10^5-9.6\cdot 10^5]$,
$\tau\in [9.6\cdot 10^5- 1.6\cdot 10^6]$, $\tau\in[4\cdot 10^6-6\cdot 10^6]$  and 
plot their cumulative distribution over their distance from the Sun in Fig.~ \ref{obser_select}. 
It is clear that for pulsars of middle ages  $\tau\in[1000, 1.6\cdot 10^6]$ the radial distributions are similar. 
The Kolmogorov-Smirnov test confirms, with a probability of more than 99\%,
that these four samples can be described by a single distribution function. 
On the other hand the radial distribution of the group of older pulsars with $\tau\in[4\cdot 10^6, 6\cdot 10^6]$ years 
is different. Therefore, our conclusions  below are correct only for
pulsars from the four first age intervals.

Let us consider a cylinder in the Galaxy, centered on the Sun, with a
radius 10 kpc and with its axis normal to the Galactic plane. 
Our selection is similar to those by~\cite{Lyne98}, but with a
larger cylinder  radius. 
As we study the relative quantities, the total number of objects which fall in the cylinder is not important. 

We can show that the observational selection which does not conceal pulsars with certain ages is harmless for
our analysis.
Let us divide our cylinder with 10 kpc radius into two parts: the inner cylinder A with radius 1.5 kpc and remained part B
of the large cylinder. The objects in cylinder A are belong to the local pulsar population.
Designate the number of pulsars with spin-down ages in the interval $[\tau, \tau+\Delta\tau]$ inside the cylinder A as $n_{\alpha}$  and 
as $n_{\beta}$ for those which are in part B of the large cylinder. For second interval of the spin-down ages 
$[\tau', \tau'+\Delta\tau]$ we designate the corresponding values as $n_{\gamma}$ in A and $n_{\delta}$. 
As it known from paper by \cite{Lyne98} the observational selection effects are weak for object in the cylinder A. 
Therefore, we can estimate the ratio $n_{\alpha}/n_{\gamma}$ with high level of confidence. 
As it was discussed above the radial distribution function for pulsars of different spin-down ages are similar. 
It means that
\begin{equation}
\frac{n_{\beta}}{n_{\alpha}}= \frac{n_{\delta}}{n_{\gamma}}
\label{relations}
\end{equation}
Gathering the pulsars from parts A and B of the large cylinder together we can write the next relation:
\begin{equation}
\frac{n_{\alpha}+n_{\beta}}{n_{\gamma}+n_{\delta}}=
\frac{n_{\alpha}(1+n_{\delta}/n_{\gamma})}{n_{\gamma}(1+n_{\delta}/n_{\gamma})}
= \frac{n_{\alpha}}{n_{\gamma}}
\label{relations_fin}
\end{equation} 
We find that the count of pulsars in the large cylinder does not change the result. 
In the other hand, the larger number of pulsars located in the large cylinder allows us to make more 
precise conclusions.

\section{Results}

A new statistical method which allows us to restore history of magnetic fields decay for middle age neutron stars was developed 
by~\cite{Igoshev2012}.
This method was applied to a real sample of radio pulsars from the ATNF catalog. We find that the surface magnetic fields 
of middle age radio pulsars decay following a modified power law:
\begin{equation}
\label{appr_res}
f(t) = \left(\left(a\frac{t}{t_0}\right)^{\gamma}+c\right)^{-1} 
\end{equation}
with parameters $\gamma=1.17$, $a=0.034$, $c=0.84$ and $t_0=10^4$ years.
It was found that observational selection is negligible when we take into consideration only part of all galactic volume. 
This effect is connected with similarity of radial distribution for radio pulsars of different ages in range $[700-1.6\cdot 10^6]$ 
years.

We also estimate the pulsar birthrate supposing that pulsars with the spin-down age over $4\cdot 10^4$ years 
have not experienced significant magnetic-field decay. This produces a
birth rate of about $2.9$ pulsars per century.

\end{document}